\documentclass[twocolumn,showpacs]{revtex4}
\usepackage{graphicx}

\newcommand{\be}{\begin{equation}}
\newcommand{\ee}{\end{equation}}
\newcommand{\ba}{\begin{eqnarray}}
\newcommand{\ea}{\end{eqnarray}}
\newcommand{\ban}{\begin{eqnarray*}}
\newcommand{\ean}{\end{eqnarray*}}

\usepackage{enumerate}

\begin{document}

\title{Identity method to study chemical fluctuations \\
in relativistic heavy-ion collisions}

\author{Marek Ga\'zdzicki}
\affiliation{Goethe-Universit\"at Frankfurt, Max-von-Laue Str. 1, 
D-60438 Frankfurt am Main, Germany}
\affiliation{Institute of Physics, Jan Kochanowski University, 
ul. \'Swi\c etokrzyska 15, PL-25-406 Kielce, Poland}

\author{Katarzyna Grebieszkow}
\affiliation{Faculty of Physics, Warsaw University of Technology,
ul. Koszykowa 75, PL-00-662 Warszawa, Poland}

\author{Maja Ma\' ckowiak}
\affiliation{Faculty of Physics, Warsaw University of Technology,
ul. Koszykowa 75, PL-00-662 Warszawa, Poland}

\author{Stanis\l aw Mr\' owczy\' nski}
\affiliation{Institute of Physics, Jan Kochanowski University, 
ul. \'Swi\c etokrzyska 15, PL-25-406 Kielce, Poland}
\affiliation{So\l tan Institute for Nuclear Studies, ul. Ho\.za 69, 
PL-00-681 Warszawa, Poland}

\date{March 15, 2011}

\begin{abstract}

Event-by-event fluctuations of the chemical composition of the hadronic final state 
of relativistic heavy-ion collisions carry valuable information on the properties
of strongly interacting 
matter produced in the collisions. However, in experiments incomplete particle
identification distorts the observed fluctuation signals. 
The effect is quantitatively studied 
and a new technique for measuring chemical fluctuations, the 
identity method, is proposed. The method fully eliminates the effect of incomplete 
particle identification. The application of the identity method to 
experimental data is explained.

\end{abstract}

\vspace{0.3cm}

\pacs{25.75.-q, 25.75.Gz}


\maketitle

\section{Introduction}

Event-by-event fluctuations of chemical (particle-type) composition of hadronic final states 
of relativistic heavy ion collisions are expected to be sensitive to properties of strongly 
interacting matter produced in the collisions~\cite{Koch:2008ia}. 
Specific fluctuations can signal the onset of deconfinement when the collision 
energy becomes sufficiently high to create droplets of 
quark-gluon plasma~\cite{Gorenstein:2003hk}.  
At higher collision energies, where the quark-gluon phase 
is abundantly produced at the early collision stage, large chemical fluctuations can 
occur as the system hits the  critical point 
of strongly interacting matter in the course 
of its temporal evolution~\cite{Koch:2005vg,Koch:2005pk}. 
It is thus certainly of interest to study event-by-event 
chemical fluctuations experimentally. First data coming from the 
CERN SPS~\cite{Afanasev:2000fu,Alt:2008ca,Kresan:2009qs} and BNL RHIC~\cite{Abelev:2009if} 
were already published. The results are not very conclusive yet and more systematic 
measurements are needed. In addition the question arises whether data analysis methods 
can be improved.

In real experiments it is impossible to determine uniquely the type of every detected 
particle. The identification requires measurements of particle electric charge and 
mass. Precise mass measurements are experimentally difficult and expensive. 
For this reason analyzes of chemical fluctuations are usually performed in a limited 
acceptance where particle identification is relatively reliable. 
However,  sensitivity to fluctuations
of range larger than the acceptance window is then  lost 
and signals from fluctuations of shorter range are usually diluted.  Furthermore 
it should be noted
that results on fluctuations, unlike those on single particle spectra, cannot be corrected 
for the limited acceptance.  Often it is possible to enlarge the acceptance, but 
at the expense of a significant contamination of the sample  by misidentified particles.
The effect of particle misidentification can distort measured fluctuations. Thus,
incomplete particle identification is a serious obstacle to the precise
measurement of chemical fluctuations. 

Although it is usually impossible to identify every detected particle, one can in general 
determine with high accuracy the percentage (averaged over many interactions) of, 
say, kaons among produced hadrons. This information will be shown to be sufficient to fully 
eliminate the effect of incomplete identification. 
In this paper we propose a new experimental 
technique called the {\em identity method} which achieves the goal independently 
of the specific properties of the chemical fluctuations under study.

In the study of chemical fluctuations
the NA49 Collaboration~\cite{Afanasev:2000fu,Alt:2008ca,Kresan:2009qs}  
used  the measure $\sigma_{\rm dyn}$, 
which is defined as the difference between fluctuations measured in real 
and {\it mixed} events.
The effect of particle misidentification is accounted for 
by including it in the  mixed events.
The STAR Collaboration~\cite{Abelev:2009if} used, in addition to the $\sigma_{\rm dyn}$
measure, the quantity  $\nu_{\rm dyn}$. The latter one assumes that particles are 
uniquely identified. 

It was suggested long ago~\cite{Gazdzicki:1997gm,Mrowczynski:1999sf} to quantify 
chemical fluctuations by the measure $\Phi$~\cite{Gazdzicki:1992ri} which proved 
to be efficient in experimental studies of event-by-event fluctuations of particle transverse 
momentum~\cite{Anticic:2003fd,Anticic:2008vb}, electric charge~\cite{Alt:2004ir}, and 
quite recently of azimuthal angle~\cite{Cetner:2010vz}. The $\Phi$ measure, unlike 
$\sigma_{\rm dyn}$ and $\nu_{\rm dyn}$,  is a strongly intensive measure of fluctuations. 
Namely, its magnitude is independent of the number and of the distribution (fluctuation) of the number 
of particle sources, if the sources are identical and independent from each other. This 
feature, which is discussed in detail in \cite{Gorenstein:2011vq}, is important in experimental 
studies of relativistic heavy-ion collisions where the collision centrality is never fully 
controllable.  However, up to now it was unclear how to correct  measurements of  
chemical $\Phi$ for the effect of particle misidentification.

The identity method, which is developed here, uses the fluctuation measure $\Psi$, 
a simple modification of $\Phi$.  The measure $\Psi$, similarly to $\Phi$, is 
strongly intensive but the modification allows to correct the measurements for 
the effect of particle misidentification. Below we show that the measure $\Psi$ 
can be factorized into a coefficient, which represents the effect of misidentification, 
and the quantity $\Psi_{\rm CI}$ which corresponds to the value $\Psi$ 
would have for complete identification. 
The misidentification coefficient can be determined from the data in a model
independent way.
Therefore, the identity method provides the value of the fluctuation measure as 
it would be obtained in an experiment in which 
every particle is uniquely identified. 

Before the identity method is presented we introduce and discuss in 
Sec.~\ref{sec-measures} the fluctuation measures $\sigma_{\rm dyn}$, $\nu_{\rm dyn}$, 
$\Phi$, and $\Psi$. In Sec.~\ref{sec-misidentification}  we demonstrate 
by a Monte Carlo simulation how the effect of misidentification distorts the chemical 
fluctuations as quantified by $\nu_{\rm dyn}$,  $\Phi$ and $\Psi$. The identity method 
is formulated in Sec.~\ref{sec-id-method}.  Instead of conclusions we present in the 
last section the steps required to apply the identity method to experimental  data.
In order to simplify the presentation, we consider chemical fluctuations of events 
composed only of kaons and pions. Clearly, kaons and pions can be replaced 
by particles of any other sort. 

\section{Measures of fluctuations}
\label{sec-measures}

As mentioned in the Introduction,  fluctuations of chemical composition of final 
states of relativistic heavy-ion collisions can be studied in several ways. 
The NA49 Collaboration~\cite{Afanasev:2000fu,Alt:2008ca,Kresan:2009qs}
measured event-by-event fluctuations of the particle ratios $K/\pi, \; K/p, \; p/\pi$
and determined the quantity $\sigma_{\rm dyn}$ defined as
\be
\sigma_{\rm dyn} = {\rm sgn}(\sigma^2_{\rm data} - \sigma^2_{\rm mixed})
 \sqrt{|\sigma^2_{\rm data} - \sigma^2_{\rm mixed}|}~,
\ee
where $\sigma_{\rm data}$ and  $\sigma_{\rm mixed}$ is the relative width 
(the width divided by the mean) of the event-by-event particle ratio distribution 
in, respectively, the 
data and  artificially generated mixed events where every particle comes from 
a different real event. The fluctuations present in mixed events are due to
the effect of particle misidentification and the 
statistical noise caused by the finite number of particles.

The STAR Collaboration used~\cite{Abelev:2009if} the quantity 
$\nu_{\rm dyn}$ to measure chemical fluctuations. For the 
case of a two-component system of pions and kaons $\nu_{\rm dyn}$ is defined as 
\be
\label{nu-dyn}
\nu_{\rm dyn} = \frac {\langle N_K (N_K-1) \rangle}  
{\langle N_K \rangle ^2} +  \frac {\langle N_\pi (N_\pi-1) \rangle}     
{\langle N_\pi \rangle ^2} - 2 \frac{\langle N_K N_\pi \rangle} 
{\langle N_K \rangle \langle N_\pi \rangle}~,
\ee 
where $N_K$  and $N_\pi$ are the numbers of kaons and pions in a given event 
and $\langle ... \rangle $ denotes averaging over events.  
The quantity $\nu_{\rm dyn}$
is defined in such a way that, in particular, 
it vanishes when the multiplicity distributions of pions 
and kaons are both poissonian ($\langle N_i (N_i -1) \rangle =
\langle N_i \rangle^2$, $i=\pi, \, K$) and independent from each other
($\langle N_K N_\pi \rangle = \langle N_K \rangle \langle N_\pi \rangle$). Thus, 
it is constructed to quantify the deviations of the fluctuations from the poissonian noise. 
For  large enough particle multiplicities one finds the approximate relation 
$\nu_{\rm dyn} \approx \sigma^2_{\rm data} - \sigma^2_{\rm mixed}$ 
which gives $\nu_{\rm dyn} \approx {\rm sgn} (\sigma_{\rm dyn}) \: \sigma^2_{\rm dyn} $ 
\cite{Abelev:2009if}.  We note that the quantity $\nu_{\rm dyn}$ implicitly assumes 
unique identification of all particles.

As already noted, it was advocated long ago~\cite{Gazdzicki:1997gm,Mrowczynski:1999sf} 
to employ the measure $\Phi$~\cite{Gazdzicki:1992ri} to study chemical fluctuations. The 
measure is defined in the following way. One introduces the variable $z \equiv x - \overline{x}$, 
where $x$ is a single particle characteristic such as the transverse momentum 
or azimuthal angle. The over-line denotes averaging over the single particle inclusive 
distribution. The event variable $Z$, which is a multi-particle analog of $z$, is defined as 
$Z  \equiv \sum_{i=1}^{N}(x_i - \overline{x})$, where the sum runs over the $N$ particles in 
a given event. By construction, $\langle Z \rangle = 0$. The measure $\Phi$ is finally
defined as
\be
\label{phi-def}
\Phi  \equiv \sqrt{\langle Z^2 \rangle \over \langle N \rangle} -
\sqrt{\overline{z^2}}~.
\ee

The measure
$\Phi$ vanishes in the absence of inter-particle correlations. This situation
is discussed in some detail below for the case of chemical fluctuations.
Here we note that the measure also possesses another important  
property - it is strongly intensive which means that it is independent of the 
number and of the distribution (fluctuation) of the number of particle sources, if the sources 
are identical and independent from each other. In particular, if a nucleus-nucleus 
collision is a simple superposition of nucleon-nucleon interactions, then 
$\Phi_{AA} = \Phi_{NN}$. The strongly intensive property is a very valuable feature of 
$\Phi$, as centrality selection in relativistic heavy-ion collisions is never 
perfect and events of different numbers of particle sources are always 
mixed up. The strongly intensive property is also desirable when different centralities 
or different colliding systems are compared. For a discussion of strongly
intensive quantities see~\cite{Gorenstein:2011vq}.

The analysis of chemical fluctuations 
can be performed with the help of 
$\Phi$ in two different but fully equivalent ways. 
In the first method~\cite{Gazdzicki:1997gm}, using the identity variable, 
chemical fluctuations are treated in analogy to fluctuations 
of transverse momentum. In the second method $\Phi$ is calculated from
the moments of the multiplicity distributions~\cite{Mrowczynski:1999sf}.

We next describe the first method for the example 
of a two-component system of pions and kaons. 
One defines the single particle variable $x$ as $x = w_K$, where $w_K$ is called 
the {\it kaon identity} and 
$w_K^i=1$ if the $i-$th particle is a kaon and  $w_K^i =0$ if the $i-$th 
particle is a pion.  This implies unique particle identification. 
One then directly uses the definition (\ref{phi-def}) to evaluate $\Phi$. 

Let us now discuss the most important case for 
which the measure $\Phi$ of chemical fluctuations 
vanishes. Since the inclusive distribution of $w_K$ equals
\be
P(w_K)  = \left\{ \begin{array}{ccl} 
\frac{\langle N_\pi \rangle}{\langle N \rangle}  & {\rm for} & w_K=0~, \\[2mm]
\frac{\langle N_K \rangle}{\langle N \rangle}  & {\rm for} & w_K=1~,
\end{array} \right. 
\ee
one finds 
\be
\overline{z^2} \equiv \overline{w_K^2} - \overline{w_K}^2 
= \frac{\langle N_K \rangle}{\langle N \rangle} 
\bigg(1 - \frac{\langle N_K \rangle}{\langle N \rangle}  \bigg)~.
\ee
When inter-particle correlations are absent, the distribution of particle 
identities in events of multiplicity $N$ reads 
\be
\label{dist-N-w}
P_N (w_K^1,w_K^2, \dots, w_K^N) = {\cal P}_N  
P(w_K^1) \, P(w_K^2) \cdots P(w_K^N)~,
\ee
where ${\cal P}_N$ is an arbitrary multiplicity distribution of particles of any 
type. One shows that $\langle Z^2 \rangle$ computed with the event distribution 
(\ref{dist-N-w}) equals $\langle Z^2 \rangle =  \langle N \rangle \overline{z^2}$ 
and consequently,  $\Phi = 0$.

In the second method the measure $\Phi$ of chemical 
fluctuations is obtained from the moments of the experimentally measured 
multiplicity distributions of kaons and pions. 
As shown in Ref.~\cite{Mrowczynski:1999sf}, one has
\ba
\label{z2bar}
\overline{z^2} 
&=& { \langle N_K \rangle \langle N_\pi \rangle \over 
\langle N \rangle^2 }~,
\\ \nonumber
{\langle Z^2 \rangle \over \langle N \rangle} 
&=&
{\langle N_\pi \rangle^2 \over \langle N \rangle^3}
\big(\langle N_K^2 \rangle - \langle N_K \rangle^2 \big)
+{\langle N_K \rangle^2 \over \langle N \rangle^3} 
\big(\langle N_\pi^2 \rangle - \langle N_\pi \rangle^2 \big) 
\\ \label{Z2/N}
&&- 2 \,{\langle N_K \rangle \langle N_\pi \rangle \over \langle N \rangle^3}
\big(\langle N_K N_\pi \rangle - \langle N_K \rangle \langle N_\pi \rangle\big)~,
\ea
which substituted in Eq.~(\ref{phi-def}) give the measure $\Phi$.  

The formulas (\ref{z2bar}, \ref{Z2/N}) clearly show that $\Phi$, like $\nu_{\rm dyn}$,  
vanishes when the multiplicity distributions of pions and kaons are both poissonian 
and independent from each other. However, more generally, $\Phi$ vanishes for any multiplicity distribution
provided it satisfies (\ref{dist-N-w}). The distribution (\ref{dist-N-w}) leads to the
multiplicity distribution of the form
\ba
{\cal P}_{N_K N_\pi} &=& {\cal P}_{N_K + N_\pi}
\\ [2mm] \nonumber
&\times& {N_K + N_\pi \choose N_K}
\bigg(\frac{\langle N_K \rangle}{\langle N \rangle} \bigg)^{N_K}
\bigg(1 - \frac{\langle N_K \rangle}{\langle N \rangle}  \bigg)^{N_\pi}~,
\ea
with the moments
\ba
\langle N_K (N_K -1) \rangle = \frac{\langle N_K \rangle^2}{\langle N \rangle^2}
\langle N (N -1) \rangle~,
\\ [2mm]
\langle N_\pi (N_\pi -1) \rangle = \frac{\langle N_\pi \rangle^2}{\langle N \rangle^2}
\langle N (N -1) \rangle~,
\\ [2mm]
\langle N_K  N_\pi \rangle 
= \frac{\langle N_K \rangle \langle N_\pi \rangle}{\langle N \rangle^2}
\langle N (N -1) \rangle~.
\ea
One checks that $\Phi$ and $\nu_{\rm dyn}$ both vanish when these moments
are substituted into Eq.~(\ref{Z2/N}) and Eq.~(\ref{nu-dyn}), respectively.

\begin{figure*}[t]
\centering
\vspace{0.4cm}
\includegraphics[width=0.49\textwidth]{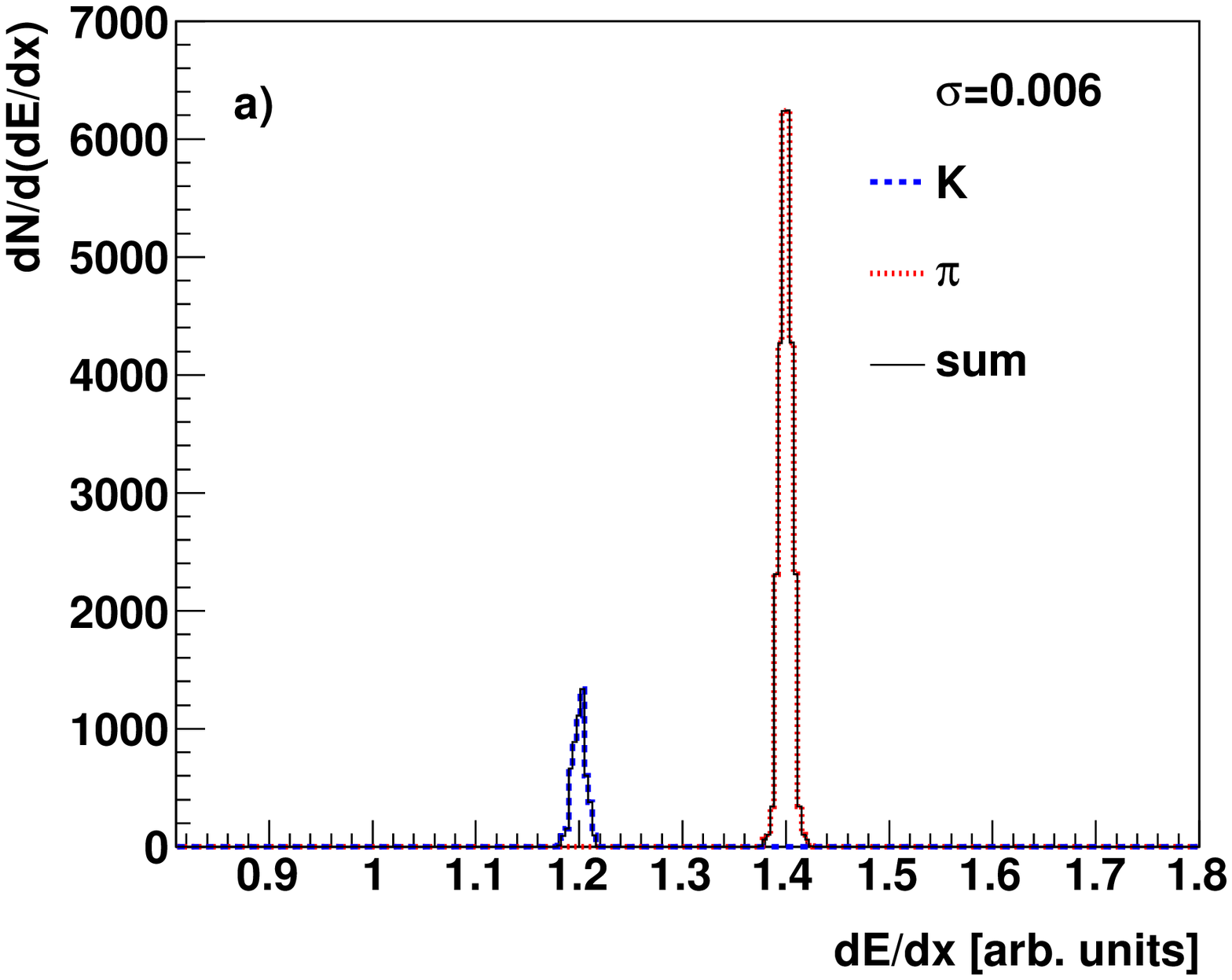}
\hspace{0.1cm}
\includegraphics[width=0.49\textwidth]{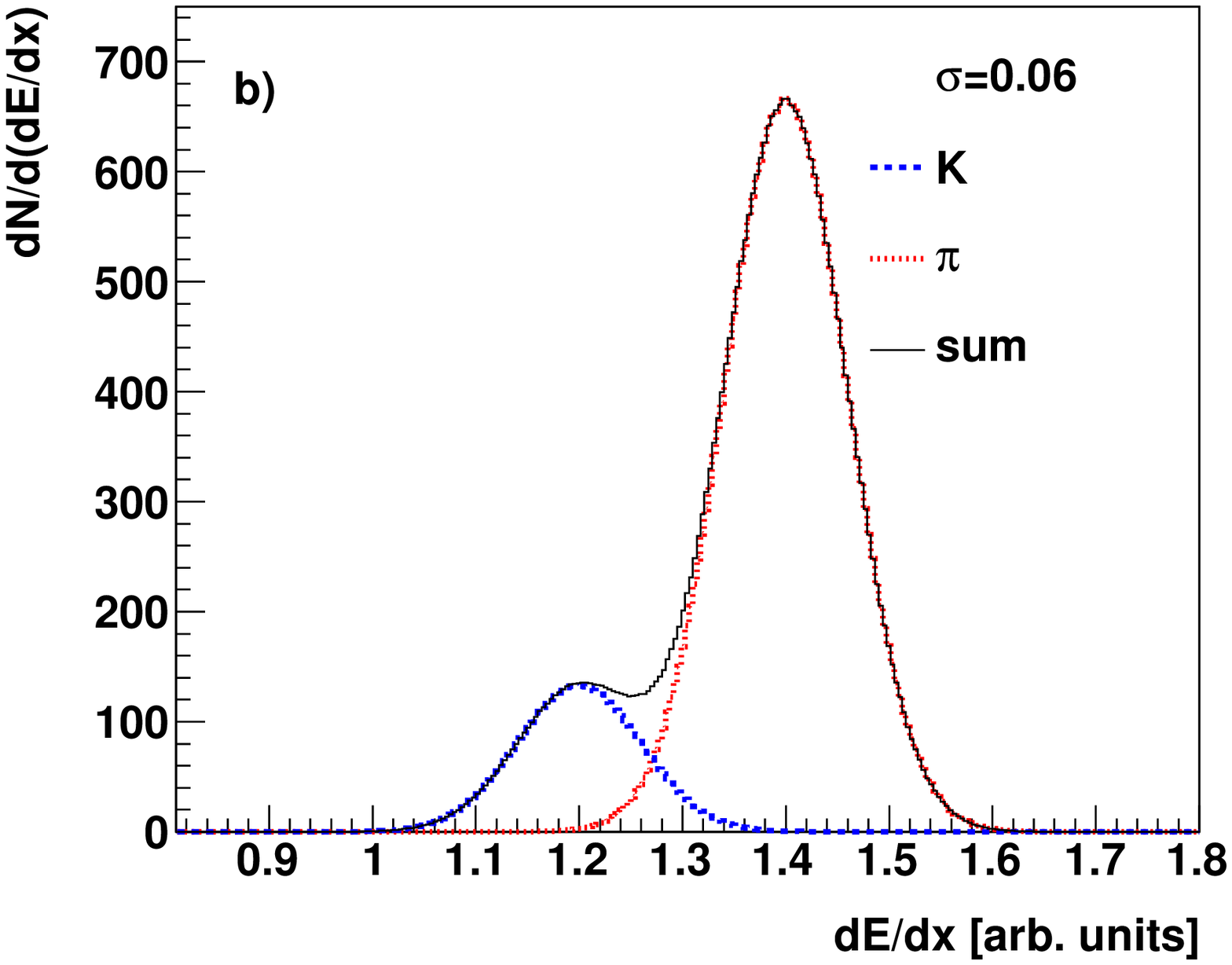}
\vspace{-0.5cm}
\caption{(Color online) The distribution of $dE/dx$ with  non-overlapping peaks 
of pions and kaons (a), which allows unique particle identification, and the 
distribution with overlapping peaks (b), which does not allow unique identification.}
\label{fig-dEdx_res}
\end{figure*} 

It appears convenient for our further considerations to modify $\Phi$ to 
the form 
\be
\label{Psi-def}
\Psi = \frac{\langle Z^2 \rangle}{\langle N \rangle}
-\overline{z^2}~,
\ee 
which preserves the properties of $\Phi$ - it vanishes in the absence of
inter-particle correlations and it is strongly intensive. The measure $\Psi$ 
will  be used to formulate the identity method for the study of chemical fluctuations. 
When expressed through moments of the multiplicity distribution, it equals
\ba
\label{Psi-moments}
\Psi 
&= &
\frac{1}{\langle N \rangle^3}
\Bigg[\langle N_\pi^2 \rangle \langle N_K \rangle^2 +
\langle N_\pi \rangle^2 \langle N_K^2 \rangle
\\ \nonumber
&-& 2 \langle N_\pi \rangle \langle N_K \rangle
\langle N_\pi N_K \rangle
- \langle N_\pi \rangle^2 \langle N_K \rangle
- \langle N_\pi \rangle \langle N_K \rangle^2
\bigg]~.
\ea
Comparing Eq.~(\ref{nu-dyn}) to Eq.~(\ref{Psi-moments}) one finds
that $\Psi$ and $\nu_{\rm dyn}$ are proportional to each other:
\be
\label{Phi-nu-dyn}
\Psi = \frac {\langle N_\pi \rangle^2 \langle N_K \rangle^2}  
{\langle N \rangle ^3} \; \nu_{\rm dyn}~.
\ee 
We note here that $\nu_{\rm dyn}$ is not intensive but it becomes
even strongly intensive when multiplied by $\langle N \rangle$, 
$\langle N_K \rangle$ or $\langle N_{\pi} \rangle$.

\section{Effect of misidentification}
\label{sec-misidentification}

As mentioned in the Introduction, complete identification of every particle is
impossible. In this section we show how the incomplete particle identification 
influences the magnitudes of fluctuation measures. For this purpose we 
considered  a simple model of chemical fluctuations where the multiplicity of
pions is poissonian with a mean value of 100 and the number of kaons 
is 20\% of the number of pions (strict correlation of the numbers of kaons
and pions). Actually, $N_K$ is taken as the integer number closest 
to $N_\pi/5$  which is smaller or equal to $N_\pi/5$. The fluctuation measures
can be easily computed analytically for the model but our aim here is to simulate
the effect of incomplete particle identification.

There are many experimental techniques to measure particle mass. We discuss 
here the effect of misidentification referring to measurements of energy loss, 
$dE/dx$, in a detector material. This method is applied by the experiments NA49, NA61 and 
STAR. The detectors are equipped with Time Projection Chambers in which $dE/dx$ 
is measured. The value of $dE/dx$ can be used to identify particles 
because it depends both on particle mass $m$ and momentum $p$ 
in the combination of velocity ($\beta = p/\sqrt{m^2 + p^2}$). 
In the case of large separation 
of the energy loss distributions of pions and kaons, as  schematically shown
in Fig.~\ref{fig-dEdx_res}a, almost unique particle identification is possible. 
This is, however, not possible when the measured  pion and kaon  
$dE/dx$ distributions overlap, as illustrated in  Fig.~\ref{fig-dEdx_res}b. 

\begin{figure*}[t]
\centering
\vspace{-0.3cm}
\includegraphics[width=0.49\textwidth]{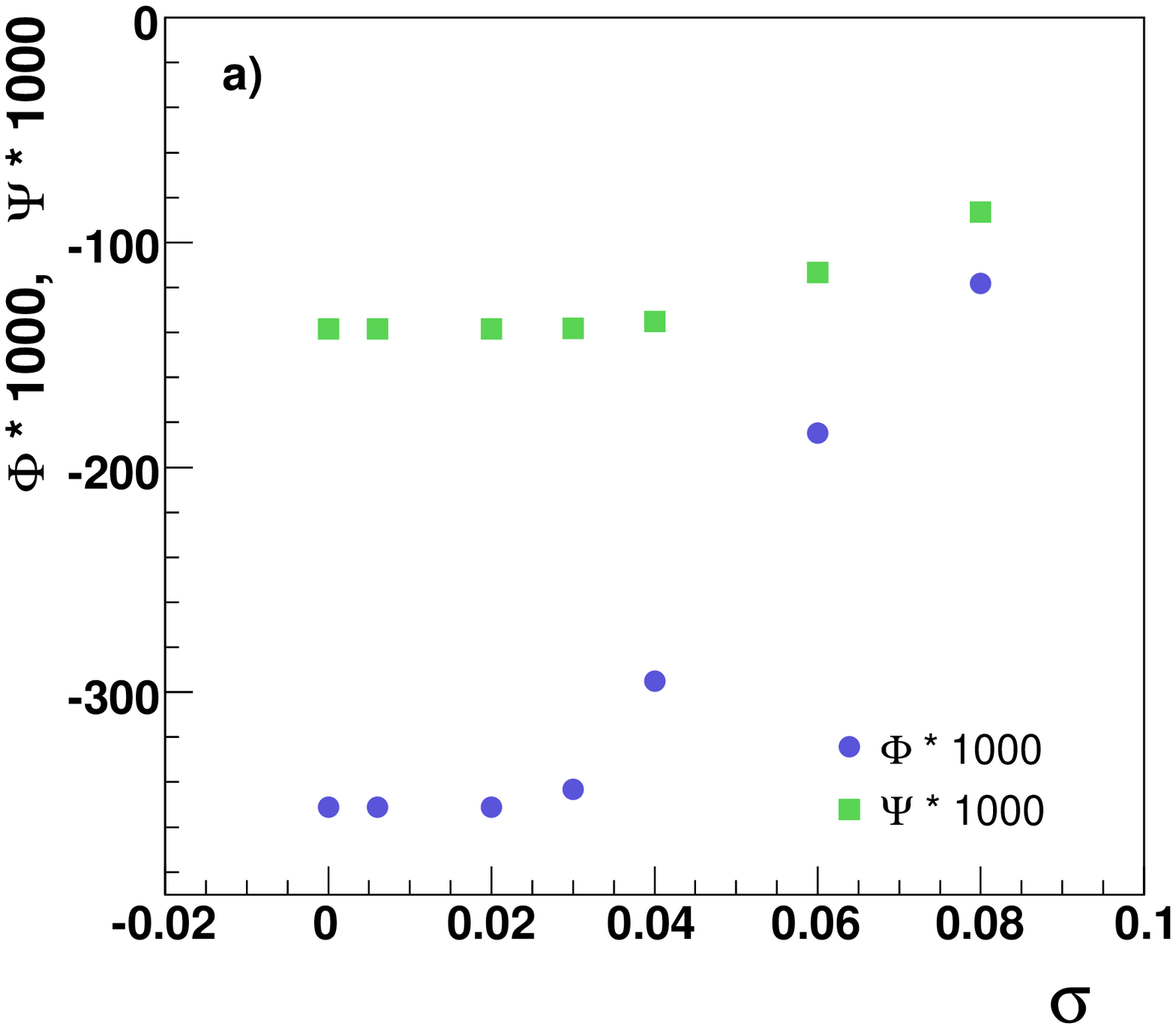}
\includegraphics[width=0.49\textwidth]{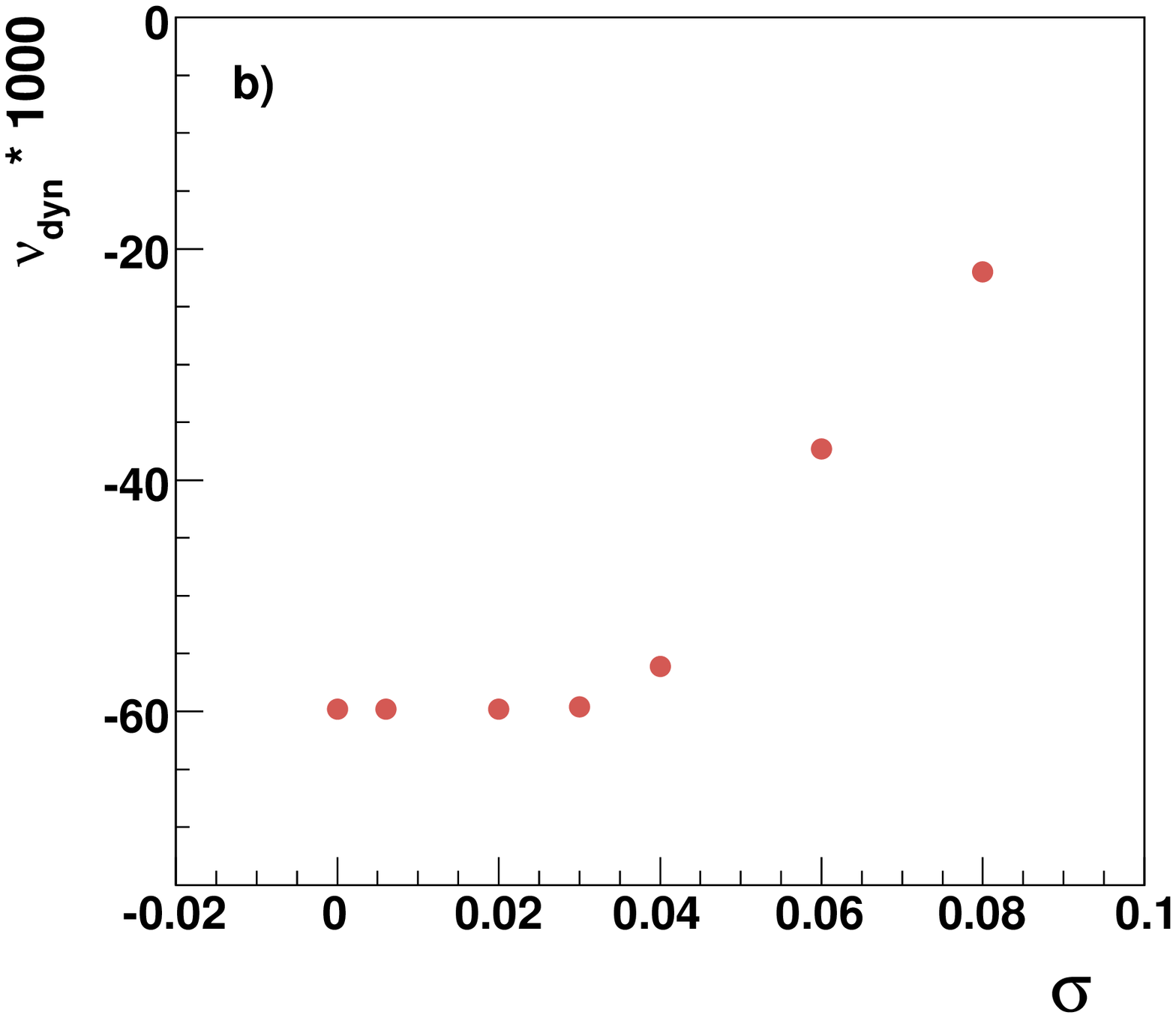}
\vspace{-1.0cm}
\caption{(Color online) The measures of chemical fluctuations $\Phi$, $\Psi$ (a) and 
$\nu_{\rm dyn}$ (b) as functions of the width of the energy-loss distribution.}
\label{fig-dEdx_res_phi_psi_nu}
\end{figure*}

Performing the Monte Carlo simulations we assumed that the  $dE/dx$ distributions 
of pions and kaons are gaussians centered, respectively, at 1.4 and  1.2 in arbitrary units. 
They  are normalized to the mean multiplicity of pions and of kaons, respectively. 
In order to quantify the bias caused by particle misidentification a simple particle
identification scheme is used, namely, a particle is identified as pion if  $dE/dx > 1.3$ 
and as kaon if  $dE/dx \le 1.3$. The width $\sigma$ of both gaussians 
is chosen to be the same but its value is varied from 0 to 0.08.  With growing  width 
of the peaks of the $dE/dx$ distribution,  the fraction of misidentified particles 
obviously increases. The results of the simulation are illustrated in 
Fig.~\ref{fig-dEdx_res_phi_psi_nu} where the fluctuation measures $\Phi$, $\Psi$ 
and $\nu_{\rm dyn}$ are shown as a function of $\sigma$. As seen, the magnitudes 
of  $\Phi$, $\Psi$ and $\nu_{\rm dyn}$ decrease as the fraction of 
misidentified particles grows and the measures vanish when particle identification 
becomes totally random. This sizable and experimentally unavoidable effect was 
the main motivation to develop the identity method which fully eliminates the problem.  

\section{Identity Method}
\label{sec-id-method}

The identity method, which is described here for a two-component system of pions and 
kaons, utilizes the measure $\Psi$ defined by Eq.~(\ref{Psi-def}). However, the kaon 
identity $w_K$ is not limited to either 1 or 0 any more, but can take
any value from the interval $[0,1]$. 

In the previous section we assumed that particles are identified according to
the energy-loss distribution. To make the presentation of the identity method 
more general we assume here that particle identification is achieved by 
measurement of particle mass not specifying the particular experimental technique
which is used for this purpose. Since any measurement is of finite resolution, 
we deal with continuous distributions of observed masses of pions and kaons
which are denoted as $\rho_\pi (m)$ and $\rho_K (m)$, respectively.  They 
are normalized as
\be
\label{norm-rho-K}
\int dm \,\rho_\pi (m) = \langle N_\pi \rangle~,
\;\;\;\;\;\;
\int dm \,\rho_K (m) = \langle N_K \rangle~.
\ee
The kaon identity is defined as:
\be
w_K (m)  \equiv \frac{\rho_K (m)}{\rho (m)}~,
\ee
where $\rho (m) \equiv \rho_\pi (m) + \rho_K (m)$, and is 
normalized as
\be
\label{norm-rho}
\int dm \,\rho (m) =  \langle N \rangle 
\equiv \langle N_\pi \rangle + \langle N_K \rangle~.
\ee

\begin{figure*}[t]
\vspace{0.3cm}
\centering
\includegraphics[width=0.49\textwidth]{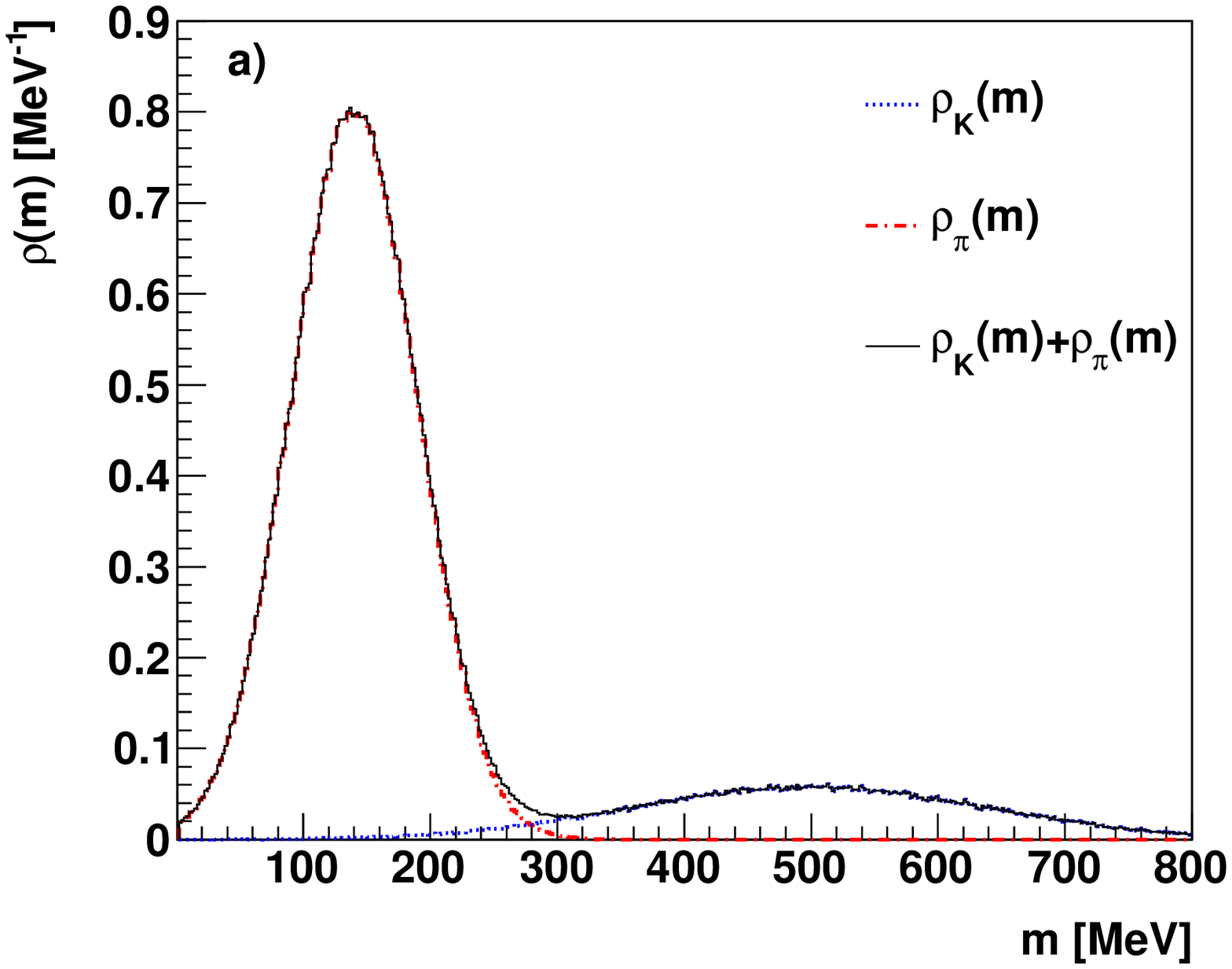}
\hspace{0.1cm}
\includegraphics[width=0.49\textwidth]{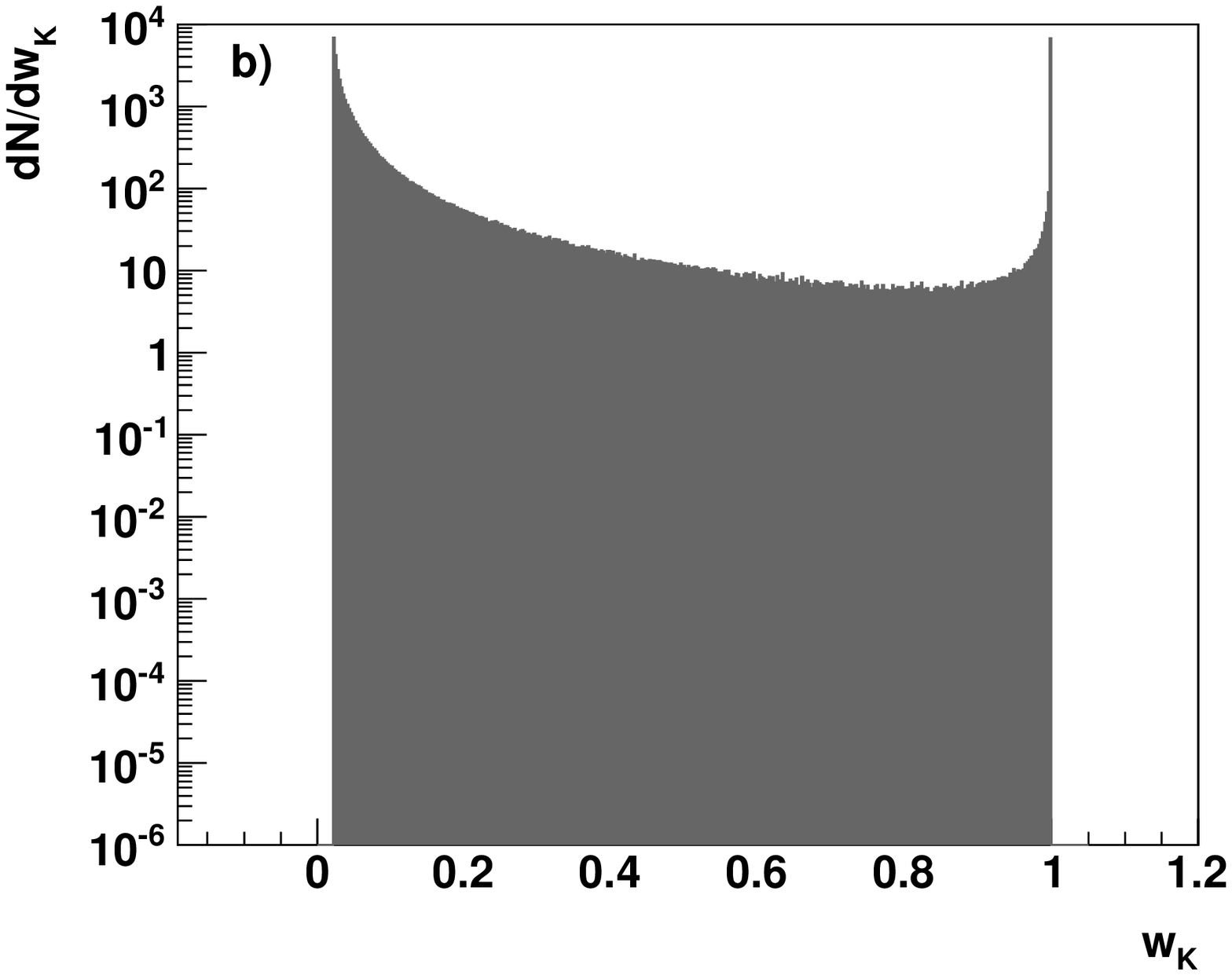}
\vspace{-0.5cm}
\caption{(Color online) The distributions of observed masses of pions $\rho_\pi (m)$, 
of kaons $\rho_K (m) $, and their sum $\rho (m)$ (a) and the corresponding 
distribution of kaon identity $w_{K}$ (b).}
\label{rho_real_identif}
\end{figure*}

If the distributions $\rho_\pi (m)$ and $\rho_K (m)$ do not overlap, 
the particles can be uniquely identified and $w_K = 0$ for a pion and  $w_K = 1$ 
for a kaon.  When the distributions $\rho_\pi (m)$ and $\rho_K (m)$ 
overlap, $w_K$ can take the value of any real number from $[0,1]$. Figure~\ref{rho_real_identif} 
illustrates the latter case. The mass distributions are shown in  
Fig.~\ref{rho_real_identif}a and the distribution of  kaon identity in 
Fig.~\ref{rho_real_identif}b.
The peaks close to~0 and~1 in Fig.~\ref{rho_real_identif}b correspond to the
mass regions in which pions, respectively kaons, are well identified.
The $w_K$ values around 0.5 correspond to particles for which
the measured mass is in the transition region between the kaon and pion
peaks ($m \approx 280$~MeV) in the distribution $\rho(m)$ 
shown in Fig.~\ref{rho_real_identif}a. 

Let us now explain how the fluctuation measure $\Psi$ is calculated once the mass 
distributions were experimentally obtained. The single particle variable entering
Eq.~(\ref{Psi-def}) is defined as  in Sec.~\ref{sec-measures}:  
$z \equiv w_K - \overline{w_K}$. The bar denotes the inclusive average 
which is computed as follows: 
\ba
\overline{w_K} &\equiv& \frac{1}{\langle N \rangle } 
\int dm \,\rho (m) \, w_K(m)
\\ \nonumber
&=& \frac{1}{\langle N \rangle } \int dm \,\rho_K (m) 
= \frac{\langle N_K \rangle }{\langle N \rangle }~.
\ea
Analogously one finds $\overline{w_K^2}$ and 
$\overline{z^2} \equiv \overline{w_K^2} - \overline{w_K}^2 $.
The quantity $\langle Z^2 \rangle$ is obtained as 
\be
\langle Z^2 \rangle = 
\frac{1}{N_{\rm ev}} \sum_{n=1}^{N_{\rm ev}}
\bigg( \sum_{i=1}^{N_n} w_K^i - N_n \, \overline{w_K}\bigg)^2~,
\ee
where $N_{\rm ev}$ is the number of events and $N_n$
is the multiplicity of the $n-$th event. Substituting  $\langle Z^2 \rangle$,
$\overline{z^2}$ and $\langle N \rangle$ into Eq.~(\ref{Psi-def}),
one finds the measure $\Psi$, the magnitude of which, however, 
is biased by the effect of particle misidentification. 
Next we discuss the correction procedure. 

As shown in the Appendix, the measure $\Psi$ can be
expressed through the moments of the multiplicity distributions of pions
and kaons as
\be
\label{Psi-final}
\Psi = A 
\bigg(\frac{\langle N_K \rangle}{\langle N \rangle}
- \overline{u_K} \bigg)^2~,
\ee
where
\ba
\nonumber
A  &\equiv&
\frac{1}{\langle N \rangle}
\bigg[\langle N_\pi^2 \rangle
\frac{\langle N_K \rangle^2 }{\langle N_\pi\rangle^2}
+ \langle N_K^2 \rangle - \langle N_K \rangle
- \frac{\langle N_K \rangle^2 }{\langle N_\pi\rangle}
\\ \label{A}
&& \;\;\;\;\;\;\;\;\;\;\;\;\;\;\;\;\;\;\;\;\;\;\;\;
-2 \langle N_\pi N_K \rangle \frac{\langle N_K \rangle}{\langle N_\pi\rangle}
\bigg]~,
\ea
and 
\be
\overline{u_K} 
\equiv \frac{1}{\langle N_K \rangle }
\int dm \,\rho_K (m) \, w_K(m)~.
\ee

In the case of complete particle identification (CI) 
the distributions $\rho_\pi (m)$ and $\rho_K (m)$ 
do not overlap and thus $\overline{u_K} = 1$. Then, the result for $\Psi$ is:
\be
\label{Psi-CI}
\Psi_{\rm CI} = 
A  \bigg(\frac{\langle N_K \rangle}{\langle N \rangle}
- 1 \bigg)^2~,
\ee
which is equivalent to the expression (\ref{Psi-moments}).

Although particle-by-particle identification is usually difficult, statistical 
identification is reliable. In the latter case, we do not know whether a given particle is a kaon, 
but we know the average numbers of kaons and of pions. We introduce the concept  
of {\em random identification} which assumes that for every particle the 
probability of being a kaon equals $\langle N_K \rangle/\langle N \rangle$.  Such a 
situation is described by mass distributions of the form
\begin{displaymath}
\rho_i(m) = \left\{
\begin{array}{ccc}
0 & {\rm for}& m < m_{\rm min}~,
\\[2mm]
\frac{\langle N_i \rangle}{ m_{\rm max} - m_{\rm min}} & {\rm for}& 
m_{\rm min} \le m \le m_{\rm max}~,
\\ [2mm]
0 & {\rm for}& m_{\rm max} < m~,
\end{array}
\right.
\end{displaymath}
where $i=\pi, K$ and $m_{\rm min}$ and $m_{\rm max}$ denote lower and upper limits
of the measured mass range, respectively. With this distribution
\be
\label{wk-randnom}
\overline{u_K} = \overline{w_K} = 
\frac{\langle N_K \rangle}{\langle N \rangle}~.
\ee
When $\overline{u_K}  = \langle N_K \rangle /\langle N \rangle$ 
is substituted into Eq.~(\ref{Psi-final}), $\Psi = 0$.
Thus, the measure $\Psi$ vanishes when 
particle identification is random. 

Finally, we arrive at the crucial point of the considerations. 
It appears that the 
measure $\Psi$ can be expressed as
\be
\label{conjecture}
\Psi = \Psi_{\rm CI} \Big(1 - \frac{V_I}{V_R} \Big)^2~,
\ee
where $\Psi_{\rm CI}$ is the measure $\Psi$ for the complete identification,
as given by Eq.~(\ref{Psi-CI}). The quantities $V_I$ and $V_R$ are the
values of the integral  
\be
\label{V}
V \equiv \int dm \, \rho(m) \, w_K(m) \, \big(1 - w_K(m) \big)~,
\ee
evaluated for the cases of imperfect and random identification, respectively.

One proves the equality (\ref{conjecture}) by observing that 
\be
\label{reso}
V_I = \langle N_K \rangle ( 1 - \overline{u_K})~,
\;\;\;\;\;\;
V_R = \frac{ \langle N_K \rangle \langle N_\pi \rangle}
{\langle N \rangle}~.
\ee
Substituting Eqs.~(\ref{Psi-CI}, \ref{reso}) into the equality (\ref{conjecture}), 
one obtains the formula (\ref{Psi-final}).  Actually, the relation (\ref{conjecture}) 
was first discovered by performing various numerical simulations and only then it was
proven analytically. Equation~(\ref{conjecture}) allows one to experimentally obtain 
$\Psi_{\rm CI}$ from $\Psi$. Thus, the effect of misidentification is fully corrected by 
the factor $(1- V_I/V_R)^2$ which measures the quality of the applied procedure of 
particle identification. The factor is independent of the correlations under study and
can be determined from experimental data.

\section{Experimental procedure}

The application of the identity method to experimental data is not difficult. We present 
here a step-by-step procedure to obtain  the measure $\Psi_{\rm CI}$ of fluctuations 
of kaons (or any other selected particle type) with 
respect to all particles 
(represented by the sum of kaons and pions in the previous sections).
It is important to note that if misidentification occurs between more
than two particle types, the identity method does not allow to study relative
fluctuations of two of them, e.g. of kaons and pions in the presence of protons.

We come back to the specific, but typical, example 
considered in Sec.~III of 
particle identification via measurements of
particle energy loss in the detector material. The energy loss
$dE/dx$ is denoted by $X$. The energy loss of particles of a given type depends 
on the particle mass and momentum (via the velocity) 
and detector characteristics. The 
distribution of $X$ is usually a multi-dimensional function, which can be determined
experimentally by averaging over particles from many interactions. The energy
loss distribution, which for a given particle momentum is typically fitted by a sum 
of four gaussians corresponding to electrons, pions, kaons and (anti-)protons, allows 
one to determine the average multiplicities of kaons and of all particles. Having
obtained this information, one should proceed as follows.

\begin{enumerate}[(i)]
\setlength{\itemsep}{1pt}

\item Extract the energy-loss distribution of kaons $\rho_K(X)$ from the inclusive 
distribution $\rho(X)$. The distributions  should be normalized as 
$$
\int dX \, \rho(X) = \langle N \rangle~, \;\;\;\;\;\;
\int dX \, \rho_K(X) = \langle N_K \rangle~.
$$

\item 
Determine the  kaon identity 
$$w_K(X)=\frac {\rho_K(X)} {\rho(X)}
$$
for every registered particle. 

\item
Compute the fluctuation measure $\Psi$ from the definition (\ref{Psi-def}). 
This is the {\em raw} value of $\Psi$ which is not corrected yet for the effect 
of misidentification.

\item
Knowing the mean multiplicities, compute the quantity
$$
V_R = \frac{\langle N_K \rangle (  \langle N \rangle-\langle N_K \rangle)} 
{\langle N \rangle}~.
$$

\item 
Using all particles calculate the integral 
$$
V_I= \int dX \rho(X) \, w_K(X) \,\big(1-w_K(X)\big)~.
$$

\item 
Determine the fluctuation measure $\Psi_{\rm CI}$, which is free of
the effect of misidentification, as 
$$
\Psi_{\rm CI} = \frac{\Psi}{(1-V_I/V_R)^2}~.
$$

\end{enumerate}

The experimental data accumulated by the NA49 Collaboration are
currently under analysis using the identity method proposed
in this paper~\cite{Mackowiak:2011sc}.

\acknowledgments

We are very grateful to Peter Seyboth for strong encouragement and
numerous critical discussions. We are also indebted to Mark Gorenstein
for critical reading of the manuscript.  This work was partially supported by 
the Polish Ministry of Science and Higher Education under grants N N202 204638 
and 667/N-CERN/2010/0 and by the German Research Foundation under 
grant GA 1480/2-1.

\begin{widetext}

\appendix*

\section{}

We express here the measure $\Psi$ through the moments of multiplicity 
distributions of pions and kaons. For this purpose one writes $\langle Z^2 \rangle$ 
as
\ba
\label{Z2-1}
\langle Z^2 \rangle &=& \sum_{N_\pi=0}^\infty \sum_{N_K=0}^\infty
{\cal P}_{ N_\pi N_K}
\int dm_1^\pi P_\pi (m_1^\pi) \int dm_2^\pi P_\pi (m_2^\pi)
\cdots \int dm_{N_{\pi}}^\pi P_\pi (m_{N_{\pi}}^\pi)
\\[2mm] \nonumber
&\times&
\int dm_1^K P_K (m_1^K) \int dm_2^K P_K (m_2^K)
\cdots \int dm_{N_K}^K P_K (m_{N_K}^K)
\\[2mm] \nonumber
&\times&
\Big(w_K(m_1^\pi) + w_K(m_2^\pi) + \dots + w_K(m_{N_\pi}^\pi)
+ w_K(m_1^K) + w_K(m_2^K) + \dots + w_K(m_{N_K}^K)
- (N_\pi + N_K) \overline{w_K} \Big)^2~,
\ea
where ${\cal P}_{ N_\pi N_K}$ is the multiplicity distribution
of pions and kaons; 
$P_\pi (m) \equiv \rho_\pi(m)/ \langle N_\pi \rangle $ and 
$P_K (m) \equiv \rho_K(m)/ \langle N_K \rangle$ are the mass 
distributions of pions and kaons, respectively. 

Equation~(\ref{Z2-1}) gives
\ba
\label{Z2-2}
\langle Z^2 \rangle &=&
\langle N_\pi \rangle \overline{u_\pi^2} + 
\langle N_K \rangle \overline{u_K^2} +
\langle N^2\rangle \overline{w_K}^2
+ \langle N_\pi (N_\pi -1) \rangle \overline{u_{\pi}}^2
+ \langle N_K (N_K -1) \rangle \overline{u_K}^2
\\[2mm] \nonumber
&+& 2\langle N_\pi N_K \rangle \overline{u_\pi} \, \overline{u_K}
- 2\langle N N_\pi \rangle \overline{w_K} \, \overline{u_\pi}
- 2\langle N N_K \rangle \overline{w_K} \, \overline{u_K}~,
\ea
where
\be
\overline{u_\pi^n} 
\equiv \frac{1}{\langle N_\pi \rangle }
\int dm \,\rho_\pi (m) \, w_K^n(m)~,
\;\;\;\;\;\;
\overline{u_K^n} 
\equiv \frac{1}{\langle N_K \rangle }
\int dm \,\rho_K (m) \, w_K^n(m)~,
\ee
with $n=1,2$. Since
\be
\frac{\langle N_\pi \rangle }{\langle N \rangle} \;
\overline{u_\pi^n} 
+ \frac{\langle N_K \rangle }{\langle N \rangle}\;
\overline{u_K^n} 
= \overline{w_K^n}~,
\ee
Equation~(\ref{Z2-2}) provides
\ba
\label{Z2-3}
\langle Z^2 \rangle =
\langle N \rangle \overline{w_K^2} 
&+& 
\Big[\langle N^2\rangle +
\langle N_\pi (N_\pi -1) \rangle 
\frac{\langle N\rangle^2}
{\langle N_\pi \rangle^2}
- 2\langle N N_\pi \rangle 
\frac{\langle N\rangle}{\langle N_\pi \rangle}
\Big]\overline{w_K}^2
\\[2mm] \nonumber
&-&
2\Big[\langle N_\pi (N_\pi -1) \rangle 
\frac{\langle N\rangle \langle N_K \rangle}
{\langle N_\pi \rangle^2}
- \langle N N_\pi \rangle 
\frac{\langle N_K\rangle}{\langle N_\pi \rangle}
+ \langle N N_K\rangle 
- \langle N_\pi N_K \rangle 
\frac{\langle N\rangle}{\langle N_\pi \rangle}
\Big] \overline{w_K} \, \overline{u_K}
\\[2mm] \nonumber
&+& 
\Big[\langle N_\pi (N_\pi -1) \rangle 
\frac{\langle N_K\rangle^2}
{\langle N_\pi \rangle^2}
+ \langle N_K(N_K-1) \rangle 
- 2\langle N_\pi N_K \rangle 
\frac{\langle N_K\rangle}{\langle N_\pi \rangle}
\Big] \overline{u_K}^2~.
\ea
Keeping in mind that $\overline{z^2} = \overline{w_K^2} - {\overline{w_K}}^2$ and 
$\overline{w_K} = \langle N_K\rangle/ \langle N \rangle$,  one finds after somewhat 
lengthy calculations the formula (\ref{Psi-final}) with $A$ given by Eq.~(\ref{A}).

\end{widetext}


\end{document}